\documentclass[apjl]{emulateapj}

\usepackage{multirow}

\slugcomment{The Astrophysical Journal Letters}

\shorttitle{X-ray emission from the super-Earth host GJ~1214}
\shortauthors{Lalitha et al.}

\begin{document}


\title{X-ray emission from the super-Earth host GJ~1214}


\author{S.~Lalitha$^{1}$, K. Poppenhaeger$^{2}$, K.P.~Singh$^{1}$, S.~Czesla$^{3}$ $\&$ J.H.M.M.~Schmitt$^{3}$ }
\affil{1. Tata Institute of Fundamental Research, Homi Bhabha road, Mumbai 400005, India\\
2. Harvard-Smithsonian Center for Astrophysics, 60 Garden St., Cambridge, 02138, MA, USA \\
3. Hamburger Sternwarte, Universit\"at Hamburg, Gojenbergsweg 112, 21029 Hamburg, Germany.}



\begin{abstract}
Stellar activity can produce large amounts of high-energy radiation, which is absorbed by 
the planetary atmosphere leading to irradiation-driven mass-loss.  
We present the detection and an investigation of high-energy emission in a transiting 
super-Earth host system, GJ~1214, 
based on an \textit{XMM-Newton} observation.
We derive an X-ray luminosity L$_X$=7.4$\times10^{25}$ 
erg s$^{-1}$ and a corresponding activity level of $\log(\frac{L_X}{L_{bol}}$)$\sim$ -5.3.  
Further, we determine a coronal temperature 
of about $\sim$3.5~MK, which is typical for coronal emission of moderately 
active low-mass stars. We estimate that GJ~1214~b evaporates at a 
rate of 1.3$\times10^{10}$~g~s$^{-1}$ and has lost a total of $\approx$2-5.6 M$_{\oplus}$.

\end{abstract}



\section{Introduction}
The discovery of the first exoplanet, 51~Peg~b, in 1995 \citep{mayor_1995}
 marked the dawn of the new field called the exoplanetary 
science. Over $\sim$1450 exoplanets have been discovered to date.
While initially most of the detected planets were Hot Jupiters, 
the frontier is inevitably approaching the Earth-mass regime, 
having already reached a large number of so-called super-Earths, i.e., planets with a mass between 1.9 and 10 
M$_\oplus$. \cite{leger_2009} reported the first transiting super-Earth system (CoRoT-7b); 
soon after, \cite{charbonneau_2009}, reported the discovery of the second transiting 
super-Earth -- namely, GJ 1214~b. 
This planet orbits a nearby (14.5 pc) M4.5-dwarf at a distance of 0.014 AU once every 1.58 days. 
GJ~1214b, has a radius of 2.68~R$_{\oplus}$ and only $\sim$ 6.55 times the Earth’s mass. 

The photometric variability detected by the MEarth project suggests the presence 
of starspots on the surface of GJ~1214. The absence of Hydrogen Balmer $\alpha$ line emission and 
the long photometric period, however, indicate a rather 
mild/low active star \citep{charbonneau_2009}. 
Nonetheless, GJ 1214 shows flaring activity and even spot crossing signatures in 
visual transit photometry \citep[see light curves in][]{kundurthy_2011}, 
clearly demonstrating that it can by no means be completely inactive.

Interestingly, the mass and radius of GJ 1214~b suggest a density of only 
$1.9 \pm 0.4$ g cm$^{−3}$ \citep{charbonneau_2009}, 
which is inconsistent with it being composed of iron and rock such as the transiting super-Earth CoRoT-7b 
\citep{leger_2009}. The density of GJ 1214~b is too low for it to be composed of even pure water ice. 
While the exact composition of the planet, GJ 1214~b, could not yet be uniquely determined by means 
of theoretical modelling, \cite{rogers_2010} state that ``the planet almost certainly 
contains a gas component''. 
In their analysis, the authors consider three different models to explain the interior 
and the gaseous atmosphere of GJ 1214~b: 
(1) a mini-Neptune that accreted and maintained a low-mass H/He layer from the primordial nebula, 
(2) a superfluid water-world with a sublimating H$_2$O envelope, or (3) a rocky planet with a 
Hydrogen dominated atmosphere formed by recent out-gassing. None of these three scenarios is 
clearly favoured by their analysis. 
Detailed calculations of GJ 1214~b's thermal evolution by \cite{nettelmann_2011} 
favour a metal-enriched envelope. The model that they consider most likely consists of an envelope of 
85\% water, mixed with Hydrogen and Helium, which resides upon a rocky core, at near-infrared 
wavelengths. The  transmission spectroscopy has, however, 
revealed a flat atmospheric spectrum \citep{berta_2012}, suggesting either an envelope with large mean molecular 
weight or the presence of a substantial cloud layer. More detailed observations have shown 
that the cloud layer interpretation is favoured by the data \citep{kreidberg_2014}.

An important ingredient in atmospheric modelling is the strength of the surrounding 
high-energy radiation field. 
Intense high-energy radiation can heat the exosphere of close-in planets to chromospheric temperatures of 
around 10000 K, which leads to planetary mass loss \citep{lammer_2003, yelle_2004, lecavelier_2004, erkaev_2007}. 
Observations of enhanced hydrogen Ly-$\alpha$ absorption during exoplanetary transits have shown that 
mass loss rates of atomic hydrogen are of the order of $\sim 10^{9} - 10^{11}$ g s$^{−1}$ for 
the three Hot Jupiters HD~209458\,b 
\citep{vidalmadjar_2003, murray_2009, linsky_2010}, HD~189733\,b \citep{lecavelier_2010} and 
WASP-12\,b \citep{fossati_2010, haswell_2012}.
Based on theoretical models by \cite{lammer_2003}, summarised in \cite{sanz_2010}, one can also derive an estimate for the total 
mass loss rate of an exoplanet based on the stellar high-energy irradiation.
In the case of an active host star such as CoRoT-2A, the planetary mass-loss rate has been estimated to be as 
high as $\sim$4.5 $\times 10^{12}$ g s$^{−1}$ \citep{schroeter_2011}. In the case of CoRoT-7b, 
which orbits a moderately active star (L$_X$ $\sim 3 \times 10^{28}$ erg s$^{−1}$, 
the mass-loss rate has been estimated to about 
$\sim 1.3\times10^{11}$ g s$^{−1}$, corresponding to an accumulated mass 
loss of 4-10~M$_{\oplus}$ \citep{poppenhaeger_2012}. 

The ongoing mass-loss has been observed in a number of hot Jovians, however, 
the low number, the tiny transit depths, and often large distances of the 
known super-Earths make the observations in this regime more challenging. 
Here, we examine the \emph{XMM-Newton} observation of super-Earth host star GJ~1214 and report 
its X-ray detection with European Photon Imaging Camera (EPIC) detectors at soft X-ray energies. 
In \S~\ref{sec:obs} we describe the observation and the data analysis, 
in \S~\ref{sec:spec} we present our results, in  \S~\ref{sec:dis} we put GJ~1214b 
and its host star in context of similar objects and summarise our findings in \S~\ref{sec:sum}.

\section{Data analysis}\label{sec:obs}
GJ~1214 was observed with \emph{XMM-Newton} for approximately 34~ks (Obs.-ID 0724380101). 
GJ~1214~b is a transiting planet and therefore, in principle, suited for atmosphere studies. 
It has recently been shown by \cite{poppenhaeger_2013} that X-ray transits of exoplanets 
can be much deeper than their optical transits due to extended atmospheric layers. 
However, our observation did not cover the transit but an orbital phase from 0.33-0.58. 
An overview of the information on the star-planet system and our observation is given in detail in 
Table~\ref{table:system}.

We only consider data taken with the EPIC, i.e., the two MOS 
and the PN detector, which were operated in the full frame mode with the medium filter. 
Those are CCD detectors that are sensitive to X-ray photons with energies between 0.2-15\,keV. 
They provide a moderate energy resolution of $\sim$100eV at energies $<2$keV. 
Low-mass stars generally display an X-ray spectrum that peaks at energies well below 2\,keV, 
and for the mildly active star like, GJ~1214, no relevant signal was collected at energies above 2\,keV at all.

\begin{table}
\begin{center}
\caption{The properties of the GJ 1214 system and the XMM-Newton observation.} 
%Stellar distance and spectral type from SIMBAD, planetary parameters from the Exoplanet Orbit Database.}
\label{table:system}
\begin{tabular}{l  l  l}
\hline\hline
\textbf{GJ 1214:} & \\ \hline 
Distance (pc)	& 14.55$\pm$0.13 & (a)\\
Spectral type	& M4.5 & (b)\\  \hline 
\textbf{GJ 1214 b:}	& \\ \hline 
Mass ($M_{Jup}$)	& 0.0204 $\pm$ 0.0031 & (b)\\
Radius ($R_{Jup}$)	& 0.239$\pm$0.011 &  (b)\\
Orbital period (d)	& 1.58040482 $\pm$ 0.00000024 & (c)\\
Transit mid-point (JD)	& 2454980.748796$\pm$ 0.000045 &  (c) \\ \hline 
\textbf{X-ray data}	& \\ \hline 
Observation start	& 2013-09-27 18:07:21\\
Observation end		& 013-09-28 03:43:42\\
Duration (ks)		& 34.4\\
Primary instrument	& XMM-Newton EPIC\\
ObsID			& 0724380101 \\
Filer			& medium \\

\tableline

\end{tabular}
\end{center}
\footnotesize{References. (a) \cite{anglada_2013}; (b) \cite{charbonneau_2009}; (c) \cite{carter_2011}.}\\
\end{table}

No useful signal from the source is present in the reflection grating spectrometer (RGS) 
data due to its much lower throughput compared to EPIC. 
The optical monitor (OM) operated in the fast imaging mode with the UVW1 filter, but 
GJ~1214 was outside the small observing window of OM. For our analysis of the 
XMM-Newton data we used the Science Analysis System (SAS) version 13.0.0 and followed 
standard routines for the data reduction. 

%%%%%%%%%%%%%%%%%%%%%%%%%%%%%%%%%%%%%%%%%%%%%%%%%%%%%%%%%%%%%%%%%%%%%%%%%%%%%%%%%%%%%%%%%%%%%%%%%%%%%%%%%%%
%\begin{figure}[!ht]
%\begin{center}
%\includegraphics[width=0.4\textwidth]{ds9_view-2.eps}
%\caption{\label{fig:img} 
%Image of the detected X-ray source from XMM-Newton’s MOS1+MOS2+PN camera in the 0.2-2.0 keV energy band.  
%The source (small circle) and the background region (large circle) are also shown.}
%along with the source free background region.}
%\end{center}
%\end{figure}

%%%%%%%%%%%%%%%%%%%%%%%%%%%%%%%%%%%%%%%%%%%%%%%%%%%%%%%%%%%%%%%%%%%%%%%%%%%%%%%%%%%%%%%%%%%%%%%%%%%%%%%%%%%%%%%%%%%%%
%

%A faint X-ray source is clearly detected close to expected position of GJ~1214. 
%
%In Figure~\ref{fig:img}, we show the soft X-ray image of GJ~1214 obtained from merged EPIC data (MOS1+MOS2+PN). 
The merged EPIC data (MOS1+MOS2+PN) shows an X-ray excess at the nominal position of GJ~1214. 
We extracted the source signal from a circular region with 
20$\arcsec$ radius centered on the nominal proper-motion corrected position of GJ~1214 from SIMBAD.  
The positional error and the offset are about 
0.3$\arcsec$ and no other X-ray source is within a radius of 1$\arcmin$ around the position of GJ~1214, 
making the identification unambiguous.

\section{Results}\label{sec:spec}
\subsection{X-ray luminosity of GJ~1214}

%The detailed source counts and background counts are given in Table~\ref{tab:counts}. 
The PN detector ran for a total of 32.76\,ks, and 97.1 net source counts with significance 
level of 5.3 $\sigma$ 
were collected in the 0.2-2\,keV band, 
yielding a PN count rate of 2.96$\pm$0.32 counts/ks. The combined two MOS detectors ran for 34.38\,ks, collecting 38.1 
net source counts with significance level of 3.4 $\sigma$ 
in the 0.2-2\,keV band. The average count rate for a single MOS detector is, therefore, 0.55$\pm$0.21 counts/ks.

We estimated the coronal temperature using the hardness ratio from the MOS1, MOS2 and PN source counts. 
Hardness ratio (HR) for the EPIC was defined as
$HR = \small{\frac{H - S}{H + S}}$, where  H is the number of counts between
0.7 and 2.0 keV (hard band) and S is the number of counts between 0.2 and 0.7 keV (soft band). 
HR determined from the EPIC data is 0.02$\pm$0.24. 

%%%%%%%%%%%%%%%%%%%%%%%%%%%%%%%%%%%%%%%%%%%%%%%%%%%%%%%%%%%%%%%%%%%%%%%%%%%%%%%%%%%%%%%%%%%%%%%%%%%%%%%%%%%
\begin{figure}
\begin{center}
\includegraphics[width=0.45\textwidth, angle=0]{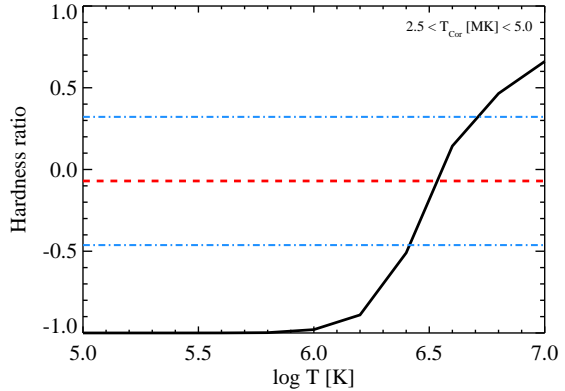}
\caption{\label{fig:hrtemp}  Dependence of mean coronal temperature on the expected hardness ratio of 
\emph{XMM-Newton's} EPIC detectors (black curve). 
The measured hardness ratio of GJ~1214 with 68$\%$ credibility is plotted as red dashed and blue dotted-dashed lines, 
respectively.}
\end{center}
\end{figure}

We calculated the theoretically predicted hardness ratios, assuming different coronal temperatures. 
We carried out this modelling using Xspec V 12.8.1 \citep{xspec}, using a coronal plasma model (APEC model)
with a single temperature component whose temperature was varied. 
Solar elemental abundances from \cite{grevesse_1998} were used.
In Figure~\ref{fig:hrtemp}, we show the results along with 
the measurements for GJ~1214. The coronal temperature of GJ~1214 is constrained to lie between 
2.5 and 5.0 MK.  

Assuming an average coronal temperature of 3.5 MK we calculated the X-ray flux of 
GJ~1214. Using WebPIMMS, we converted the mean count rate of \emph{XMM-Newton} PN detector 
into an X-ray flux of 3.1$\pm$1.9$\times$10$^{-15}$ erg~s$^{-1}$~cm$^{-2}$ 
in the 0.2-2.0 keV energy band. The average signal from the MOS detectors yields a flux 
estimate of 3.2$\pm$1.3$\times$10$^{-15}$ erg~s$^{-1}$~cm$^{-2}$ and is in 
line with the result from the PN detector. 
Given a distance of $\sim$14.55$\pm$0.13~pc, the observed X-ray flux in PN translates into a luminosity of 
$7.8\pm4.8\times10^{25}$ erg~s$^{-1}$.
The bolometric luminosity of GJ~1214 can be calculated from 
 $L_{bol}=10^{0.4(4.8-m_v-bc+5log(d)-5)}L_{\odot}$,
 with m$_v$ denoting the apparent visual magnitude of the star, bc denoting the bolometric correction, 
 d the distance in pc 
 and L$_{\odot}$ the solar bolometric luminosity. 
 Given the apparent V magnitude of 14.67 and K magnitude of 8.78, 
 the bolometric corrections are determined 
 using \cite{worthey_2011}; and the bolometric luminosity is found to be  
 1.4 $\times$10$^{31}$ erg s$^{-1}$. 
 Comparing the X-ray luminosity (L$_X$) to the bolometric luminosity (L$_{bol}$), we determine the 
 ratio of $log(\frac{L_X}{L_{bol}})$$\approx$ -5.3. The X-ray luminosity as well as  
the ratio of luminosities point to a mildly active star. 

\subsection{X-ray variability of GJ~1214}
\begin{figure}
\begin{center}
\includegraphics[width=0.47\textwidth, angle=0]{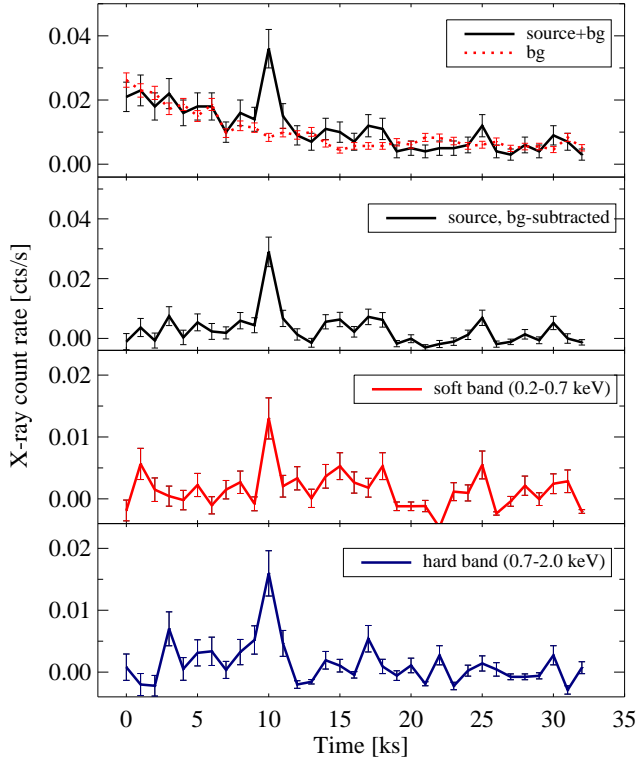}
%\vspace{-20.0pt}
\caption{\label{fig:lightcurve}  (a) Soft X-ray light curve extracted from the source 
region of GJ~1214 (solid) and the adjacent background region (red dotted), (b) 
the background-subtracted X-ray light curve of GJ~1214 (0.2-2.0\,keV), (c) and (d)  
show the light curve in the soft band (0.2-0.7\,keV) and the 
hard band (0.7-2\,keV), respectively. 
In panel (b), (c) and (d) a flare-like feature in X-ray flux is 
observed at $\sim$10\,ks after the start of the observation.}
\end{center}
\end{figure}

We have used the signal from the PN camera, which is more sensitive than the MOS cameras, 
to extract X-ray light curves of GJ~1214 and for the background region for comparison, 
using a $20^{\prime\prime}$ extraction radius for the source, and a $50^{\prime\prime}$ 
extraction radius for the background.% (see Figure~\ref{fig:img}). %We used a time binning 
Data were binned every 1000 seconds in energy band of 0.2-2\,keV. The resulting light curves for the 
source and the background are shown in Figure~\ref{fig:lightcurve} (a).

%%%%%%%%%%%%%%%%%%%%%%%%%%%%%%%%%%%%%%%%%%%%%%%%%%%%%%%%%%%%%%%%%%%%%%%%%%%%%%%%%%%%%%%%%%%%%%%%%%%%%%%%%%%

The background is not constant, it shows a decline over the duration of the observation. 
We plot the background subtracted X-ray light curve in Fig.~\ref{fig:lightcurve} (b).
After subtraction of the background from the source region signal, we find that GJ~1214 
displays a sudden increase in X-ray flux about 10\,ks after the start of the observation. 
This is most likely the signature of a small flare event in GJ~1214's corona. 
A flare-like signature 
on GJ 1214 has also been observed in the r-band, where 
an energy release of 1.8$\times10^{28}$ erg has been estimated \cite{kundurthy_2011}. 

To investigate the nature of this flare further, we have split the signal from GJ~1214's 
PN light curve into two energy bands. In Fig.~\ref{fig:lightcurve} (c) and Fig.~\ref{fig:lightcurve} 
(d), we plot a soft 
X-ray light curve in 0.2-0.7\,keV energy band 
 and a hard X-ray light curve in 0.7-2\,keV energy band, respectively. 
Note that the flare signal is 
present in both bands with roughly the same amplitude. This implies that the flare did 
not cause a strong heating of the local corona, because then we would expect to see a 
stronger signature at harder X-ray energies emitted by the heated plasma.

\section{Discussion}\label{sec:dis}

\subsection{GJ~1214~b in comparison to the evaporating Jupiter HD 209458~b}

X-ray observations show that GJ 1214 b's host star is mildly active, however, the close proximity of the planet 
to the star places GJ 1214 b in an intense high-energy radiation field. It is, therefore, 
possible that GJ 1214~b is undergoing atmospheric evaporation. The energy range that is 
thought to be mainly responsible for atmospheric mass loss is the X-ray and extreme UV 
range \citep{lammer_2003}. 

\cite{sanz_2011} computed EUV spectra for several stars 
 using emission measure distribution analyses of X-ray spectra. 
 Later, \cite{claire_2012} and \cite{linsky_2014} 
 compared predicted EUV flux ratios by \cite{sanz_2011} for a few stars with the 
ratios of EUV fluxes measured in different wavelength bands. They found that the 
values are comparable in  all wavelength bands.
Hence, we extrapolated the measured X-ray luminosity to the full X-ray and extreme UV (XUV) range 
using the scaling relation of \cite{sanz_2011}:
\begin{equation}
\log L_{EUV} = (0.860 \pm 0.073) log L_X + (4.80 \pm 1.99)
\end{equation}
\begin{equation}
\log L_{XUV} = \log (L_X + L_{EUV})
\end{equation}

For GJ 1214, we find an XUV luminosity of $\log L_{XUV} = 27.09\pm0.02$\,erg\,s$^{-1}$. 
To put the irradiation of GJ 1214 b into context, we compare it with the current 
XUV irradiation of the Earth, and the XUV irradiation of the Hot Jupiter HD 209458 b. 
In Table~\ref{tab:comp}, we list the various properties of planet hosts and the 
resultant irradiation levels 
for the Earth, HD~209458~b and GJ~1214~b.
The atmosphere of HD~209458 has been observed in the UV with a very large transit 
depth \citep{vidalmadjar_2003}. The interpretations in the literature agree that a 
very extended atmosphere must be present \citep{benjaffel_2007, vidalmadjar_2003}, 
and active atmospheric escape is likely occurring because the atmosphere is extended 
beyond the Roche-lobe \citep{vidalmadjar_2004}, even if the accuracy of velocity 
measurements of this escaping atmosphere have been debated \citep{benjaffel_2008}. 
HD 209458 has been observed in X-rays, but not been detected, though, a firm upper limit of 
$\log L_X < 26.12$\,erg\,s$^{-1}$ has been established \citep{sanz_2010}.

In comparison to HD 209458 b the XUV flux 
at the planetary orbit of GJ 1214 b is substantially larger, by about an order of magnitude (Table~\ref{tab:comp}). 
We also show the respective values for the Earth, using a solar X-ray luminosity of 
$\log L_X = 27$\,erg\,s$^{-1}$ \citep{judge_2003}.

The strong high-energy irradiation of GJ 1214 b suggests that the atmosphere of this 
planet may be evaporating. We proceed by estimating how much mass the planet may be 
losing, given analytical theoretical models.

\begin{table}
\begin{center}
\caption{\label{tab:comp}Properties of the Sun, HD 209458, and GJ 1214.}
\begin{tabular}{lrrr}
\tableline
\tableline
Parameters 	 & Sun & HD 209458 & GJ~1214 \\
\tableline
Sp. type &	G2V   &	G0V		& M4.5\\
log L$_X$& 27.0&$<$26.12 &25.87\\
log $\frac{L_X}{L_{bol}}$&$\approx$-6.6 &$<$-7.6 &-5.3\\
% T$_{Cor} [MK]$& & &3.5\\

\tableline
a$_P$ (AU)&1.00 &0.047 &0.014\\
$P_{orb}$ (d)& 365	& 3.52	&  1.58\\
M$_P$ (M$_E$)& 1.00 & 219 & 6.47\\
F$_X$ $[erg~s^{-1}~cm^{-2}]$ &0.35 &$<$21.4 &135.6\\
at planet orbit& & &\\
F$_{XUV}$ $[erg~s^{-1}~cm^{-2}]$ &4.1 &$<$312 &2150\\
at planet orbit& & &\\

\tableline

\end{tabular}

\end{center}
\end{table}

\subsection{Estimated evaporation of GJ~1214b}
In the picture of energy-limited hydrodynamic mass-loss \citep{watson_1981, lammer_2003, sanz_2010},
the mass-loss rate, $\dot{M}$, amounts to:
\begin{equation}
 \dot{M} = \frac{\pi R_p^3 \epsilon F_{XUV}}{GKM_p} \, ,
 \label{eq:mdot}
\end{equation} 
where $R_p$ is the planetary radius, $F_{XUV}$ is the incident X-ray and
EUV-flux, $\epsilon=0.4$ is the heating efficiency as suggested by \cite{valencia_2010}, 
$G$ is the gravitational
constant, $M_p$ the mass of the planet, and $K$ is a parameter accounting for
Roche-lobe filling, which we assume to be unity here.  
According to \cite{valencia_2010}, the 
above expression remains valid for strongly irradiated rocky planets, 
because the atmosphere may be replenished by sublimation faster than it erodes.
Therefore, assuming a 
density of $\approx$1.9~g~cm$^{-3}$ \citep{charbonneau_2009} and substituting the XUV luminosity 
of $\log L_{XUV} \approx 27.09$\,erg\,s$^{-1}$ 
into Eq.~\ref{eq:mdot}, yields an estimate of 1.3$\times10^{10}$~g~s$^{-1}$ for mass-loss rate 
of GJ~1214~b.

The high-energy emission for super-Earth host star has been measured, so far, for only 
CoRoT-7, a subsolar mass star with spectral type G8-K0 \citep{poppenhaeger_2012}. 
They, estimated 
a mass loss rate of 0.5-1.0$\times10^{11}$~g~s$^{-1}$ for CoRoT-7b, comparable within an order of 
magnitude to the value obtained here. 

%%%%%%%%%%%%%%%%%%%%%%%%%%%%%%%%%%%%%%%%%%%%%%%%%%%%%%%%%%%%%%%%%%%%%%%%%%%%%%%%%%%%%%%%%%%%%%%%%%%%%%%%%%%
\begin{figure}
\begin{center}
\includegraphics[width=0.45 \textwidth, angle=0]{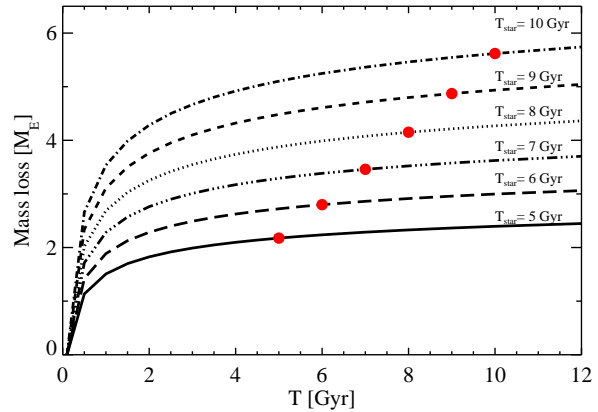}
\caption{\label{fig:loss}  Evolution of the total planetary mass-loss for the current X-ray luminosity. 
The filled points indicate the current age. }
\end{center}
\end{figure}

The stellar X-ray and UV emission is not constant over time. Young stars usually 
display much higher X-ray luminosities 
than older stars. Scaling relations for X-ray luminosity and age have been given by several authors, 
for example \cite{ribas_2005, lammer_2009}.
We assume here that the X-ray luminosity increases with age by a 
factor $(\frac{\tau}{\tau_{\star}})^{−1.23}$, 
where $\tau$ is the young stellar age when the stellar activity remains at 
a constant level (0.1 Gyr) and $\tau_{\star}$ is the current stellar age in Gyr \citep{ribas_2005}. 

Estimating the current age of the host star GJ 1214 is not an easy task. Typical age-activity 
relationships display substantial scatter, especially for older stars \citep{mamajek_2008}. 
\cite{poppenhaeger_2014} have recently compiled X-ray luminosities of old disk and halo 
dwarfs of different spectral types in order to test for exoplanetary influences on stellar 
rotation and activity. They give a typical X-ray luminosity range of $\log L_X = 25.76$ to 
$27.21$\,erg\,s$^{-1}$ for old disk/halo M dwarfs. GJ 1214 falls well into this range with 
its X-ray luminosity of $\log L_X = 25.87$\,erg\,s$^{-1}$. An age of 5-10\,Gyr 
therefore seems likely for this star. 
Inserting the time variable XUV flux into Eq.~\ref{eq:mdot} and integrating over stellar ages of 5-10\,Gyr, 
we plot the mass loss history of GJ~1214~b in Figure~\ref{fig:loss}; each curve represents 
a different age of GJ~1214 system. Considering the large uncertainty in the age of GJ~1214, 
we estimate a total mass loss of 2-5.6 M$_{\oplus}$.

\section{Summary}\label{sec:sum}
We report X-ray emission from a nearby super-Earth host, GJ~1214. In the quiescent state 
GJ~1214 is detected at soft X-ray energies (0.2-2.0~keV) with an X-ray luminosity of $\sim$7.4$\times10^{25}$erg~s$^{-1}$. This 
leads to an X-ray activity level of log $\frac{L_X}{L_{bol}}$$\approx$-5.3, combined with a likely coronal 
temperature $\approx$3.5~MK, points to a mildly active star. We notice a small flare-like signature in the X-ray light curve, 
however, we found no evidence of this event triggering any strong heating of the local corona.

Extrapolation of the X-ray luminosity to extreme UV range leads to XUV flux of 2150 erg~s$^{-1}$cm$^{-2}$ at the planetary 
orbit and a mass loss 
rate of 1.3$\times10^{10}$~g~s$^{-1}$ for GJ~1214~b. 
Integration of the stellar activity history, leads to a mass loss of $\approx$2-5.6 M$_{\oplus}$  
over the lifetime of the planet. 
 
\bibliographystyle{apj}
\bibliography{paper}

\end{document}